\documentclass[onecolumn,showpacs,preprintnumbers,amsmath,amssymb,11pt]{revtex4}
\date{March 2010}
\usepackage{epsfig}
\usepackage{latexsym}
\usepackage{amssymb}
\usepackage{booktabs}
\usepackage{graphicx}
\newcommand{\be}{\begin{equation}}
\newcommand{\ee}{\end{equation}}
\newcommand{\ba}{\begin{eqnarray}}
\newcommand{\ea}{\end{eqnarray}}
\newcommand{\bi}{\begin{itemize}}
\newcommand{\ei}{\end{itemize}}

\newcommand{\ud}{\,\mathrm{d}}

\newcommand{\half}{{\textstyle\frac{1}{2}}}

\newcommand{\quarter}{{\textstyle\frac{1}{4}}}

\newcommand{\<}{\langle}
\renewcommand{\>}{\rangle}
\newcommand{\eq}{Eq.~}

\newcommand{\la}{\label}

\newcommand{\txts}{\textstyle}

\newcommand{\dBdloga}{\frac{d\beta}{d\log a}}

\newcommand{\bx}{\boldsymbol{x}}

\newcommand{\as}{a_{\sigma}}
\newcommand{\at}{a_{\tau}}
\newcommand{\betas}{\beta_{\sigma}}
\newcommand{\betat}{\beta_{\tau}}

\newcommand{\dBsdlogas}{\frac{\partial\beta_{\sigma}}{\partial\log a_{\sigma}}}
\newcommand{\dBtdlogas}{\frac{\partial\beta_{\tau}}{\partial\log a_{\sigma}}}
\newcommand{\dBsdlogxi}{\frac{\partial\beta_{\sigma}}{\partial\log \xi}}
\newcommand{\dBtdlogxi}{\frac{\partial\beta_{\tau}}{\partial\log \xi}}

\newcommand{\dFdBs}{\frac{\partial F}{\beta_\sigma}}
\newcommand{\dFdBt}{\frac{\partial F}{\beta_\tau}}

\newcommand{\Ss}{S_{\sigma}}
\newcommand{\St}{S_{\tau}}

\newcommand{\Ns}{N_{\sigma}}

\newcommand{\Zps}{Z^{+}_{\sigma}}
\newcommand{\Zpt}{Z^{+}_{\tau}}
\newcommand{\Zms}{Z^{-}_{\sigma}}
\newcommand{\Zmt}{Z^{-}_{\tau}}

\newcommand{\bp}{\boldsymbol{p}}

\begin{document}
\title{Lattice Gauge Theory Sum Rule for the Shear Channel}

\author{Harvey~B.~Meyer}
\affiliation{Johannes Gutenberg-Universit\"at Mainz,
Institut f\"ur Kernphysik, D-55099 Mainz, Germany}
\email{meyerh@kph.uni-mainz.de}

\date{\today}

\begin{abstract}
An exact expression is derived 
for the $(\omega,\bp)=0$ thermal correlator
of shear stress in SU($N_c$) lattice gauge theory.
I remove a logarithmic divergence by taking 
a suitable linear combination of the 
shear correlator and the correlator of the energy density.
The operator product expansion shows 
that the same linear combination 
has a finite limit when $\omega\to\infty$.
It follows that the vacuum-subtracted
shear spectral function vanishes 
at large frequencies at least as fast as
$\alpha_s^2(\omega)$ and obeys a sum rule.
The trace anomaly makes a potential contribution 
to the spectral sum rule which remains 
to be fully calculated, but which I estimate
to be numerically small for $T\gtrsim 3T_c$.
By contrast with the bulk channel, the shear channel
spectral density is then overall enhanced as compared 
to the spectral density \emph{in vacuo}.
\end{abstract}

\maketitle

\section{Introduction}

The shear viscosity $\eta$ is a quantity
that universally characterizes the relaxation 
of a fluid towards equilibrium.
Indeed $\eta c^2/[k_B(e+p)]$ is the diffusion coefficient
of transverse momentum, and this process
is in many situations occurring in Nature
the dominant mechanism of dissipation
($e$ is the energy density and $p$
the pressure of the system).
The dimensionless quantity $\eta/(\hbar s)$ 
can be thought of as the ratio of a transport
time scale $\eta/[k_B(e+p)]$ 
to the thermal time scale $\hbar /(k_BT)$
and is therefore a good measure of the ability
of a fluid to flow~\cite{Kovtun:2004de,Csernai:2006zz}.

In the past years, the phenomenology 
of the RHIC heavy ion experiments 
has provided an upper bound on the ratio $\eta/s$
of hot quark matter (see~\cite{Teaney:2009qa} for a review; from here on
we set $\hbar$, $c$ and $k_B$ to unity).
In parallel to this,
the thermodynamic properties of QCD in the range 
of temperatures $100\lesssim T/{\rm MeV}\lesssim700$ 
are the subject of ongoing Monte-Carlo simulations
on a space-time lattice~\cite{Aoki:2009sc,Bazavov:2009zn,Cheng:2009zi}. 
In this computational approach,
access to the near-equilibrium properties, 
such as the shear viscosity, is limited
because lattice QCD employs the Euclidean 
formulation of thermal field theory.
Real-time properties can thus only be determined 
by analytic continuation, 
a numerically ill-posed 
problem~(see e.g.$~$\cite{Cuniberti:2001hm}, \cite{Bazavov:2009us} section 5).

Through the Kubo formula, the shear viscosity is related 
to the low-frequency part of the spectral density, 
$\eta = \pi \lim_{\omega\to0}\rho_{12,12}/\omega$
(\cite{Son:2007vk} and Refs.~therein).
Here and in the following we
denote by $\pi\rho_{\mu\nu,\rho\sigma}$ the 
imaginary part of the retarded correlator of the 
energy-momentum tensor components 
$T_{\mu\nu}$ and $T_{\rho\sigma}$.
Some constraints on the thermal behavior of the spectral 
function have been obtained by computing the Euclidean 
correlator of the stress-energy tensor 
on the lattice~\cite{Meyer:2007ic,Meyer:2007dy}.
Because the spectral density is related by an integral
transform to the Euclidean correlator, one is automatically 
led to studying the spectral density over the whole 
semi-axis of frequencies.
For that reason, it is helpful to have global 
constraints on the spectral densities, such as 
sum rules~\cite{Kharzeev:2007wb,Karsch:2007jc}. 
Very recently, I determined some of the gross features
of the bulk-channel spectral density by combining
Euclidean correlation functions with 
a spectral sum rule~\cite{Meyer:2010ii}.
The goal of this paper is to take the steps necessary to 
apply the same strategy to the shear channel.

Romatschke and Son~\cite{Romatschke:2009ng} obtained 
a shear sum rule for conformal field theories,
and proposed a modified sum rule that takes into account 
the trace anomaly, $T_{\mu\mu}$.
Here I give a derivation of the shear sum rule 
for the SU($N_c$) gauge theory in the lattice regularization.
In the continuum, the shear sum rule reads
\be
\int_{-\infty}^\infty \frac{\ud\omega}{\omega}\,
\left[\rho_{12,12}(\omega,{\bp=0},T) 
    - \rho_{12,12}(\omega,{\bp=0},0) \right] 
= 
\frac{2}{3} e(T)
-\lim_{\omega\to\infty}\Delta G(\omega,T)\,,
\la{eq:conti-sr}
\ee
where 
$\Delta G(\omega,T)$ is defined by
\eq(\ref{eq:defDG}); I will show by explicit calculation 
that its $\omega\to\infty$ limit is a finite quantity,
and that it is proportional to $e-3p$. 
The right-hand side of  \eq(\ref{eq:conti-sr})
is therefore consistent with the form of the sum rule 
given by Romatschke and Son.
Although the calculation of the coefficient remains to be completed,
I estimate its practical importance for the determination 
of the shear viscosity in the range of temperatures 
explored at the LHC and find 
it to represent at most a $5\%$ correction to 
the energy-density term on the right-hand side 
of \eq(\ref{eq:conti-sr}). As a consequence,
the right-hand side is positive at those temperatures, 
indicating that $\rho(\omega)/\omega$ is overall enhanced 
relative to the vacuum spectral density. 
This is in contrast with the bulk sum rule,
which indicates an overall depletion of the bulk 
spectral weight throughout the deconfined phase.

I start by deriving an expression for the 
$\omega=\boldsymbol{p}=0$ Euclidean shear-channel correlator
in section (\ref{sec:lsr}). 
Next I convert this identity into a spectral sum rule
for the corresponding spectral density (Sec. \ref{sec:csr}).
The contact term that has to be subtracted before using 
the spectral representation is then shown to be finite 
in section (\ref{sec:ctt}). I show that part of the finite 
term can be obtained from existing results, 
while the other remains to be calculated. 
The numerical importance of the contact term is
estimated on the basis of this partial result.
I make some final comments in section (\ref{sec:comm}).
\section{Derivation of the Lattice Sum Rule\la{sec:lsr}}
I start by introducing the essential notation, which follows 
closely~\cite{Meyer:2007fc}. The reader is referred to that 
paper for unexplained notation.
The stress tensor in the SU($N_c$) gauge theory
reads $T_{\mu\nu} = \theta_{\mu\nu} + \quarter \delta_{\mu\nu} \theta $,
\be
\theta(x) \equiv  \beta(g)/(2g) ~ F_{\rho\sigma}^a(x)  F_{\rho\sigma}^a(x)  
\qquad\qquad
\theta_{\mu\nu}(x) \equiv 
{\txts\frac{1}{4}}\delta_{\mu\nu}F_{\rho\sigma}^a F_{\rho\sigma}^a
   - F_{\mu\alpha}^a F_{\nu\alpha}^a .
\ee
The beta-function is defined by 
$qdg/dq=\beta(g)=-g^3(b_0+\dots)$
and $b_0=11N_c/(3(4\pi)^2)$,
$b_1= 34N_c^2/(3(4\pi)^4)$   in the SU($N_c$) pure gauge theory.
If $\<\dots\>_T$ denotes the thermal average at temperature $T$,
\be
\epsilon-3P = \<\,\theta\,\>_T -\<\,\theta\,\>_0\equiv \<\,\theta\,\>_{T-0}, 
  \qquad\qquad
\epsilon+P = {\txts\frac{4}{3}}  \<\,\theta_{00}\,\>_{T}  ~.\la{eq:basic}  
\ee
I will be considering the Wilson discretization of the SU($N_c$) gauge theory 
on an anisotropic lattice,
\be
S_{\rm g} =   \sum_x \betas \Ss(x)  + \betat \St(x)\,,
\ee
where $\Ss$ and $\St$ are respectively the sum of spatial
and temporal plaquettes.
The bare gauge coupling is given by $g_0^2=2N_c/\sqrt{\betas\betat}$.
The time direction is discretized more finely by a factor $\xi$,
called the (renormalized) anisotropy, than the spatial directions. 
I use the following discretizations,
\ba
\xi^{-3}~\Theta(x) &=& \Zps(\betas,\betat) \Ss + \Zpt(\betas,\betat) \St  \, \\
\xi^{-3}~\Theta_{00}(x) &=& \Zms(\betas,\betat) \Ss - \Zmt(\betas,\betat)\St  \,.
\ea
The  factor $Z_{\sigma,\tau}^\pm$ are such that 
 $\<\sum_x \Theta_{00}(x)\>{\rightarrow} 
\<\int \ud^4x \, \theta_{00}(x)\>$ in the continuum limit.

In the earlier publication~\cite{Meyer:2007fc}, I derived lattice sum rules 
for the ($\omega,\bp)=0$ two-point functions of $\theta$ and $\theta_{00}$.
In these sum rules, the functions of bare lattice parameters
$\lambda_{00}^+$ and $\lambda_{00}^-$ appear, which were defined as
\be
\lambda^{\pm}_{00}(\betas,\betat) \equiv 
\frac{1}{2}
\Big(\frac{\partial}{\partial\betas}
- \frac{\partial}{\partial\betat}\Big) (\Zms \pm \Zmt).
\la{eq:l00}
\ee
In the appendix of~\cite{Meyer:2007fc},
I gave the values of the latter in the limit $g_0\to0$,
corresponding to a lattice spacing exponentially smaller
than the confinement scale. These values are wrong.
The derivation of these values was based on the 
erroneous idea that the 
operator $\theta_{00}$ does not mix with the unit operator
\emph{even on the anisotropic lattice}.
I presently correct this mistake.
\subsection{Perturbative determination of $\lambda_{00}^\pm$}
The parameters $(\betas,\betat)$ appearing in the lattice action
are related to the renormalized parameters $(\as,\xi)$ 
in a well-defined manner (up to O($\as^2$) ambiguities; 
the spatial lattice spacing $\as$ 
is measured in units of a renormalized quantity such as a glueball mass).
It is convenient to exchange the `bare' variables $(\betas,\betat)$
for $(\beta,\xi_0)$, where 
\be
\xi_0\equiv \sqrt{\betat/\betas},\qquad \mathrm{and}\qquad
\beta\equiv\sqrt{\betas\betat}\,.
\ee
The advantage of the new variables is that
at the classical level, there is no difference between 
the bare anisotropy $\xi_0$ and the renormalized anisotropy $\xi$. 
Since the spatial and temporal plaquettes respectively yield
$\as^4 \boldsymbol{B}^2$ and $\as^2\at^2 \boldsymbol{E}^2$,
it is not hard to guess the form of $Z_{\sigma,\tau}^-$,
\ba
\Zms &=& \frac{\beta}{\xi_0^4}= \frac{\betas^{5/2}}{\betat^{3/2}}\,,
\\
\Zmt &=& \frac{\beta}{\xi_0^2}= \frac{\betas^{3/2}}{\betat^{1/2}}\,,
\ea
One then easily finds 
\ba
\lambda_{00}^+  &=& 3 + {\rm O}(g_0^2)\,,
\la{eq:l00+basic}
\\
\lambda_{00}^-  &=& 1+ {\rm O}(g_0^2)\,.
\la{eq:l00-basic}
\ea

To determine the quantum corrections, a little more work is 
needed. First, it was shown in~\cite{Meyer:2007fc} that at the isotropic point,
the renormalization factors $Z_{\sigma,\tau}^{-}$ are given by 
the renormalization of the anisotropy,
\be
\Zms\stackrel{\xi=1}{=}\Zmt\stackrel{\xi=1}{=} Z(g_0) 
= \left[\frac{\partial\xi_0(\as,\xi)}{\partial\xi}\right]_{\xi=1}\,.
\ee
The factors $\lambda_{00}^{\pm}$ evaluated at the isotropic point 
can be written as 
\ba
\la{eq:l00-}
\lambda_{00}^{-} &\stackrel{\xi=1}{=}& 
Z(g_0) 
+ \frac{1}{\beta Z} \left[ \frac{\partial^2 \beta(\as,\xi)}{\partial(\log\xi)^2}
     - \frac{1}{16} \frac{d^2\beta(\as,\xi)}{d(\log a)^2}\right]_{\xi=1}\,,
\\
\la{eq:l00+}
\lambda_{00}^{+} &\stackrel{\xi=1}{=}& 2+Z(g_0)
-\frac{1}{Z(g_0)}\left[\frac{\partial^2\xi_0(\as,\xi)}{\partial\xi^2}\right]_{\xi=1}
+\frac{g_0^4}{2Z(g_0)}\frac{dg_0^{-2}}{d\log a}\frac{dZ(g_0)}{dg_0^2}\,.
\ea
The relation between the bare and renormalized anistropy is known
to one-loop order~\cite{Karsch:1982ve}.
Indeed, two functions $c_{\sigma,\tau}(\xi)$ were computed 
there (\cite{Karsch:1982ve}, Eq. (2.24)--(2.25)), 
from which the coefficients
$c_{\sigma,\tau}'\equiv \frac{dc_{\sigma,\tau}}{d\xi}|_{\xi=1}$
and 
$c_{\sigma,\tau}''\equiv \frac{d^2c_{\sigma,\tau}}{d\xi^2}|_{\xi=1}$ 
can be obtained. The derivatives of the bare with respect to 
the renormalized anisotropy can be expressed in terms of these coefficients,
\ba
Z(g_0)=\left[\frac{d\xi_0}{d\xi}\right]_{\xi=1} &=&  1 - {\txts\frac{1}{2}}
(c_\sigma'-c_\tau') g_0^2 +{\rm O}(g_0^4)\,,
\\ 
\left[\frac{d^2\xi_0}{d\xi^2}\right]_{\xi=1} &=&
-g_0^2\left[c_\sigma'-c_\tau' + {\txts\frac{1}{2}}(c_\sigma''-c_\tau'')\right]
+{\rm O}(g_0^4)\,.
\ea
For instance, numerically~\cite{Karsch:1982ve}
\be
c_\sigma'-c_\tau' =
{\txts\frac{N_c^2-1}{N_c}}\cdot 0.146711 - N_c \cdot 0.019228\,.
\ee
Inspecting \eq(\ref{eq:l00-}), we see that the square bracket is 
at least O($g_0^2$), and therefore
\be
\lambda_{00}^{-}(g_0) = Z(g_0) +{\rm O}(g_0^4)
=  1 - {\txts\frac{1}{2}}
(c_\sigma'-c_\tau') g_0^2 +{\rm O}(g_0^4)\,. 
\la{eq:l00-g2}
\ee
Similarly,
\be
\lambda_{00}^{+}(g_0) = 
3 + \frac{g_0^2}{2}\left[c_\sigma''-c_\tau'' + c_\sigma' - c_\tau'\right]
+{\rm O}(g_0^4)\,.
\la{eq:l00+g2}
\ee
These results will be used in sections (\ref{sec:csr}) and 
(\ref{sec:ctt}).
\subsection{Sum Rule for $\theta_{11}$}
In order to be sensitive to correlations of shear stress,
we need to derive a sum rule for (say) the two-point function of 
the operator $\theta_{11}$.
For that purpose we will need to work on an anisotropic lattice, and 
will take the isotropic limit at the end.
To get at the $\theta_{11}$ correlator,
the anisotropy has to be in the $\hat1$-direction.

Since it is conventional
to associate the time-direction with the direction 
where the lattice spacing is smaller by a factor $\xi$,
I temporarily interchange the labels of the 
$x$-direction $\hat1$ and the time-direction $\hat0$.
I thus consider a lattice of dimensions 
$\infty_0\times( \Ns\times \infty_2\times \infty_3)$
and calculate the correlator of $\theta_{00}$
in this new coordinate system.
At the end I will restore the normal labels, at what point 
the direction with $\Ns$ lattice points will play the role 
of the Matsubara cycle.

The procedure now closely follows~\cite{Meyer:2007fc}.
First, consider a generic renormalization-group
invariant quantity $f(\as,\xi,T)$, which is obtained as 
the continuum limit of a function of the bare parameters,
$F(\betas,\betat,\Ns)$.
Expressing the independence of $f(\as,\xi,T)$ on $\as$ and $\xi$, 
I obtain respectively
\be
L\frac{\partial f}{\partial L} 
\left( \begin{array}{c}  1 \\ 
                         0  \end{array}\right)
= \left(  \begin{array}{c@{\quad}c} \dBsdlogas & \dBtdlogas \\
                                    \dBsdlogxi &  \dBtdlogxi \end{array}\right)
 \left(  \begin{array}{c} \dFdBs \\
                         \dFdBt    \end{array}\right)            
\ee
At the symmetric point $\xi=1$, the determinant of the matrix 
is $2\beta Z(\beta) \dBdloga$. By taking a suitable linear combination,
one finds
\be
-\frac{1}{4} L \frac{\partial f}{\partial L} = \beta Z(\beta)
\left(\dFdBs - \dFdBt \right)\,.
\la{eq:lsr-gen}
\ee
This equation holds up to O($\as^2$) effects.

I now apply \eq(\ref{eq:lsr-gen}) to the case of 
\ba
f(L) \equiv L^4 \<\theta_{00}\>\,,
\ea
which is the continuum limit of the lattice expectation value
\ba
F(\beta_\sigma,\beta_\tau,\Ns) = \Ns^4 \xi \,
\left(\<\Theta_{00}\>_{L}-\<\Theta_{00}\>_{(L=\infty)}\right)\,.
\ea
Since we are on the anisotropic lattice, it is necessary
to perform the subtraction on the infinite lattice ($L=\infty$) 
in order to remove the mixing with the unit operator.
After manipulations entirely similar to those performed 
in~\cite{Meyer:2007fc}, I get 
\be
\la{eq:lsr-th00}
\beta Z(\beta) \left(\frac{\partial F}{\partial\beta_\sigma}
 - \frac{\partial F}{\partial\beta_\tau}\right) 
= -4F +\Ns^4 \lambda_{00}^{-}\beta Z(\beta)\<S_{+}\>^{L}_{\infty}
 +\Ns^4 \lambda_{00}^{+} \<\Theta_{00}\>^{L}_{\infty}
-\Ns^4 \sum_x \<\Theta_{00}(x)\Theta_{00}(0)\>^{L}_{\infty}\,.
\ee

Returning to the normal coordinate system, where the 
compact direction is time and $f$ is given by $-\frac{1}{4}(e+p)$
at the temperature $T=1/L$, the lattice sum rule (\ref{eq:lsr-gen}) 
can be written as 
\be
a^{-4}\< {\txts\sum_x}\,\Theta_{11}(x)\Theta_{11}(0)\>_{T-0}
- \frac{\beta Z(\beta)\lambda^{-}_{00}(\beta)}{\dBdloga}(e-3p)
=(1-{\txts\frac{1}{4}}\lambda_{00}^+)(e+p) + 
 {\txts\frac{1}{16}} T^5 \partial_T \frac{e+p}{T^4}.
\ee
For completeness I reproduce the sum rule for  
the $\theta_{00}$ correlator obtained in~\cite{Meyer:2007fc},
\be
a^{-4}\< {\txts\sum_x}\,\Theta_{00}(x)\Theta_{00}(0)\>_{T-0} 
-  \frac{\beta Z(\beta)\lambda^{-}_{00}(\beta)}{\dBdloga} (e-3p)
= {\txts\frac{3}{4}} \lambda^{+}_{00}(\beta) (\epsilon+P) 
+ {\txts\left(\frac{3}{4}\right)^2} T^5 \partial_T \frac{\epsilon+P}{T^4}.
\la{eq:an_sr_m}
\ee
It is noteworthy that the difference of two-point functions 
\ba
a^{-4}{\txts\sum_x}\Big\<{\txts\frac{3}{4}}\,\Theta_{11}(x)\Theta_{11}(0) 
&\!\!-\!\!&{\txts\frac{3}{4}}\, \Theta_{00}(x)\Theta_{00}(0)\Big\>_{T-0}
\nonumber
\\
&\equiv& a^{-4}{\txts\sum_x}\Big\<{\txts\frac{1}{4}}(\Theta_{11}-\Theta_{22})(x)(\Theta_{11}-\Theta_{22})(0)
-{\txts\frac{2}{3}} \Theta_{00}(x)\Theta_{00}(0)\Big\>
\la{eq:1step}
\\
&=& {\txts\frac{3}{4}}(1-\lambda_{00}^{+}) (e+p) -{\txts\frac{3}{8}} T^5 \partial_T \frac{e+p}{T^4}
\la{eq:lsr}
\ea
is UV-finite. In writing \eq(\ref{eq:1step}) I have used the elementary identity
\be
{\txts\sum_x}\,\<\Theta_{00}(x) \Theta_{00}(0)\> =
 {\txts\sum_x}\,\<3\,\Theta_{11}(x)\Theta_{11}(0)
+ 6\,\Theta_{11}\Theta_{22}(0)\>\,.
\ee
A potential application of \eq(\ref{eq:lsr}) is a determination of the 
renormalization factor $Z(g_0)$, since it appears quadratically on the left-hand side
and only linearly on the right-hand side of the equation.
However it is probably in the continuum limit that this relation is most useful,
as described in the next section.
\section{Sum Rule in the Continuum 
and Dispersion Relation\la{sec:csr}}
We are now ready to write a sum rule for the correlator of shear stress,
$T_{12}$.
I define 
\be \la{eq:defDG}
\Delta G(\omega,T) \equiv 
\int \ud^4x\, e^{i\omega x_0} \Big\<{\txts\frac{1}{4}}
(T_{11}(x)-T_{22}(x))(T_{11}(0)-T_{22}(0))
-  {\txts\frac{2}{3}} T_{00}(x)\,T_{00}(0)\Big\>_{T-0}.
\ee
The lattice sum rule (\ref{eq:lsr}) shows that $\Delta G(0,T)$ is finite.
In the next section, I will show that also 
$\lim_{\omega\to\infty}\Delta G(\omega,T)$ is UV-finite.
I therefore write the dispersion relation for this correlator as 
(for the general method of deriving sum rules, 
see~\cite{Kapusta:1993hq,Romatschke:2009ng})
\ba 
\Delta G(0,T) -\lim_{\omega\to\infty}\Delta G(\omega,T)
&=& \int_{-\infty}^\infty \frac{\ud\omega}{\omega}\, 
\left({\txts\frac{1}{2}}\Delta\rho_{11,11}(\omega,T) 
      -{\txts\frac{1}{2}}\Delta\rho_{11,22}(\omega,T) 
-   {\txts\frac{2}{3}}\Delta\rho_{00,00}(\omega,T)\right)
\nonumber\\
&=& \int_{-\infty}^\infty \frac{\ud\omega}{\omega}\,
      \Delta\rho_{12,12}(\omega,T)   - {\txts\frac{2}{3}}Tc_v\,.
\la{eq:drln}
\ea
In the last step, I have used rotation symmetry, which implies 
$\rho_{12,12}= \half \rho_{11,11}-\half \rho_{11,22}$
(the spatial momentum $\bp$ is set to zero throughout this section).
I have also used the fact that 
$\rho_{00,00}(\omega,T) = Tc_v\omega\delta(\omega)$.

On the other hand, $\Delta G(0)$ can be rewritten identically as 
\be
\Delta G(0,T) = 
\int \ud^4x\,\Big\<
{\txts\frac{3}{4}}\theta_{11}(x)\theta_{11}(0) 
- {\txts\frac{3}{4}}\theta_{00}(x)\theta_{00}(0)
-{\txts\frac{1}{24}} \theta(x)\theta(0) 
-{\txts\frac{1}{3}} \theta_{00}(x)\theta(0)\Big\>_{T-0}\,.
\ee
In view of \eq(\ref{eq:lsr}) and in view of the (continuum) 
sum rules~\cite{Meyer:2007fc}
\ba
\< {\txts\int} \ud^4x \,\theta(x) \theta(0) \>_{T-0} 
&=& T^5 \partial_T \frac{e-3p}{T^4}
\la{eq:src1}
\\
\< {\txts\int} \ud^4x\, \theta(x)  \theta_{00}(0) \>_T &=& 
{\txts\frac{3}{4}} T^5\partial_T \frac{e+p}{T^4}\,,
\la{eq:src2}
\ea
we have, in infinite volume and in the continuum limit,
\be
\Delta G(0,T)= \frac{2}{3} \left( e
 -  Tc_v\right)\,.
\la{eq:DG0}
\ee
To reach this expression,
we have used the thermodynamic relations $c_v=\frac{\partial e}{\partial T}$, 
$s=\frac{\partial p}{\partial T}$,
and $Ts=e+p$, as well as \eq(\ref{eq:l00+basic}). Now combining \eq(\ref{eq:drln}) and (\ref{eq:DG0}),
we reach \eq(\ref{eq:conti-sr}) by recalling that, in infinite volume,
\be
\int \ud^3x\, \Big\<{\txts\frac{1}{4}} (T_{11}(x)-T_{22}(x))(T_{11}(0)-T_{22}(0))\Big\>
= \int \ud^3x\, \<T_{12}(x)\, T_{12}(0)\>\,.
\ee

In the free theory, we can check the coefficient of the 
$e=\frac{d_A\pi^2}{15}T^4$ 
term in \eq(\ref{eq:conti-sr}) ($d_A\equiv N_c^2-1$). Using
\be
\rho^{\rm free}_{12,12}(\omega,T) = \frac{d_A}{10(4\pi)^2} \frac{\omega^4}{\tanh \omega/4T}
+  \left({\txts \frac{2\pi}{15}}\right)^2 \, d_A T^4\omega\delta(\omega)\,,
\ee
the coefficient of the energy density on the
right-hand side of the sum rule is correctly reproduced.
\section{Contact terms and asymptotics of the spectral density\la{sec:ctt}}
%
The spectral sum rule (\ref{eq:conti-sr}) contains the 
quantity $\lim_{\omega\to\infty}\Delta G(\omega,T)$.
In this section we study the UV contact terms that appear in 
$\Delta G(\omega,T)$, since those are the only ones that survive 
in the infinite frequency limit. We are working in infinite volume
and start by considering the contribution of the energy density 
correlator. At finite separation in time, this correlator has 
a simple time-independent expression,
\be
\int \ud^3x\, \<T_{00}(x)\, T_{00}(0)\>
= T^2c_v\qquad \forall x_0\neq 0\,.
\ee
The lattice sum rules, on the other hand, 
determine the correlator of $T_{00}$ integrated over all times.
Combining these two pieces of information, one finds
the contact term (see~\cite{Meyer:2007fc}, section 3.3),
\be
\lim_{\omega\to\infty} \int \ud^4x\, e^{i\omega x_0}
  \<T_{00}(x)\, T_{00}(0)\>_{T-0}
=\left(-\frac{1}{2b_0g_0^2}+{\rm finite}\right) (e-3p)
-\frac{3}{4} (e+p)\,,
\ee
where equation (\ref{eq:l00-basic}) has been used.

We now come to the contribution of the shear-stress correlator 
to $\Delta G$. 
At large frequencies, the operator-product expansion (OPE)
is applicable.  The leading-order result
is~\cite{CaronHuot:2009ns}\footnote{I have checked the results
using the methods of~\cite{Novikov:1983gd}.}
\be
\int \ud^4x\,  e^{i\omega x_0} \<T_{12}(x)T_{12}(0)\>_{T-0}
\stackrel{\omega\to\infty}{\sim} 
-\frac{1}{3b_0g_0^2} (e-3p)
-\frac{1}{2}(e+p) \,.
\la{eq:T12OPE}
\ee
Thus we see that 
\be
\Delta G(\omega,T) \stackrel{\omega\to\infty}{\sim} 
 {\rm O}(1)\cdot (e-3p)\,.
\ee
In particular, $\Delta G(\omega,T)$ is finite 
in the $\omega\to\infty$ limit, and its value comes entirely
from the contribution of the $T_{\mu\mu}$ operator.

The fact that both the zero-frequency correlator $\Delta G(0,T)$ 
and the contact term $\lim_{\omega\to\infty}\Delta G(\omega,T)$
are finite implies that the correlator in coordinate space
(i.e.~as a function of Euclidean time $t$) diverges at short distances
at most as 
\be
\Delta G(t,T)\sim \frac{\alpha_s^2(1/t)}{t}\cdot
\left(
\textrm{linear combination of}~e,\,p
\right)
\ee
at small $t$. From the spectral representation of this correlator
\be
\Delta G(t,T) \stackrel{t\to 0}{\sim}
 \int_\Omega^\infty \ud\omega\, 
\left(\Delta\rho_{12,12}(\omega,T)
 - {\txts\frac{2}{3}}\Delta\rho_{00,00}(\omega,T)\right)\, e^{-\omega t}\,,
\ee
it follows that the leading behavior of the shear spectral function 
is at most
\be
\Delta\rho_{12,12}(\omega,T)\stackrel{\omega\to\infty}{\sim}
\alpha_s^2(\omega)\cdot \left(
\textrm{linear combination of}~e,\,p
\right)\,.
\ee

The leading order OPE result~(\ref{eq:T12OPE})
does not allow us to obtain 
the finite part of the Wilson coefficient of the operator
$T_{\mu\mu}$ in the OPE of 
$\lim_{\omega\to\infty}\Delta G(\omega,T)$.
However, the lattice sum rules allow us to obtain
the finite part of the $\<T_{\mu\mu}\>_{T-0}=e-3p$ coefficient in 
the OPE of the energy 
correlator~\footnote{The large-frequency limit is taken 
in the regime $\omega\ll\frac{\pi}{a}$. Equivalently, 
the contact term is defined via a separation distance scale $d$,
with $a\ll d\ll T^{-1}$, by the difference 
$(\int_0^{1/T} - \int_{d}^{1/T-d})\ud x_0\int \ud^3\bx\,\<T_{00}(x)T_{00}(0)\>_{T-0}$.
For instance, one could scale $d$ as $(aT)^{\frac{1}{2}}$.},
\be
\lim_{\omega\to\infty} \int \ud^4x\, 
e^{i\omega x_0}\<T_{00}(x)\, T_{00}(0)\>_{T-0}
= c_{\theta} (e-3p) - \left({\txts\frac{3}{4}}+{\rm O}(g_0^2)\right)(e+p).
\ee
Indeed, the coefficient takes the value~\cite{Meyer:2007fc}
\be
c_\theta = 
\frac{\lambda_{00}^{-}(g_0)Z(g_0)}{g_0^2 \frac{dg_0^{-2}}{d\log a}}
            - \frac{1}{4} + {\rm O}(g_0^4)
=  -\frac{1}{2b_0g_0^2} +\frac{1}{2b_0}    (c_\sigma' - c_\tau')
+\frac{b_1}{2b_0^2} - \frac{1}{4} + {\rm O}(g_0^2)\,,
\la{eq:cth}
\ee
where \eq(\ref{eq:l00-g2}) has been used  in the second equality.
As shown above, the leading term,
which diverges logarithmically in the lattice spacing,
cancels in $\Delta G(\omega,T)$.
The O($g_0^2$) terms vanish in the continuum limit.
For $N_c=3$, the finite terms in \eq(\ref{eq:cth}) numerically 
amount to, respectively, 
\be
\lim_{g_0\to0}\left(c_\theta +\frac{1}{2b_0g_0^2}\right)
= 2.3942 + 0.4215 - 0.2500\,.
\ee
The dominant contribution comes from the regularization-dependent 
$(c_\sigma' - c_\tau')$ term. It is associated with the lack 
of continuous translation invariance on the lattice.
We expect the regularization
dependence to cancel in $\Delta G(\omega,T)$, since it is a finite
correlator. Therefore we take the two other terms to be 
representative of the size of $T_{\mu\mu}$'s Wilson coefficient
in the OPE of $\Delta G(\omega,T)$ (after the inclusion of a factor
$\frac{2}{3}$, see \eq(\ref{eq:defDG})).
In the SU(3) gauge theory for $T>2.5T_c$, 
$(e-3p)/T^4\lesssim 0.5$ while $e/T^4\approx4.5$
is very weakly temperature-dependent~\cite{Boyd:1996bx}. 
From the size of the Wilson coefficient $c_\theta$, 
we therefore expect 
$\lim_{\omega\to\infty}\Delta G(\omega,T)$ 
in the shear sum rule \eq(\ref{eq:conti-sr})
to represent at most a $5\%$ correction to the right-hand side
of the equation.
\section{Final Comments\la{sec:comm}}
It is hoped that the spectral sum rule (\ref{eq:conti-sr})
will be useful in constraining the spectral density
at finite temperature, especially when combined 
with lattice Monte-Carlo data on 
the Euclidean correlator.
In order to be operational at all temperatures,
the remaining contribution to $\Delta G(\omega,T)$ should 
be computed in the large-frequency limit.
Since we have seen that one contribution to $\Delta G$ has
a dependence on the regularization, it is important
to compute $\Delta G$ as a whole in the same regularization.
Dimensional regularization is then probably the 
computationally most economic choice.

The sum rule (\ref{eq:conti-sr}) 
is particularly useful in the pure gauge theory 
because the mass gap in the tensor channel is very large, 
$m_{2}/T_c\approx 7.9$~\cite{Meyer:2004jc,Chen:2005mg}%
\footnote{Or $m_{2}\approx2.4$GeV
if one sets the scale with the 
Sommer parameter $r_0^{-1}=0.41$GeV}.
Thus even at $T=3T_c$, the substracted spectral density
appearing in (\ref{eq:conti-sr}) coincides 
with the spectral density up to $\omega\approx 2.6T$,
in particular the latter is unaffected by the subtraction
in the range of frequencies that determine the transport properties.
Moreover, the contribution of the lightest tensor glueball
to $\rho_{12,12}(\omega,\boldsymbol{0},T=0)$ has been calculated 
on the lattice~\cite{Meyer:2008tr,Chen:2005mg}. It is 
parametrized by a matrix element $F_T$ in the notation 
of~\cite{Meyer:2008tr}.
At $3T_c$, this contribution to the spectral 
sum rule (\ref{eq:conti-sr}) is small,
\be
\frac{2}{\left(\frac{2e(3T_c)}{3}\right)}
\int_{m_2-\epsilon}^{m_2+\epsilon} \frac{\ud\omega}{\omega}
 \rho_{12,12}(\omega,\boldsymbol{0},0)
=  
\frac{2F_T^2}{m_2\left(\frac{2e(3T_c)}{3}\right)} \approx 2\%.
\ee

At high temperatures, 
there is another contribution to the spectral sum rule 
that can be clearly isolated and identified.
For a weakly coupled field theory, 
where the spectral density admits a transport peak 
at the origin of width much smaller than temperature,
Teaney derived a sum rule~\cite{Teaney:2006nc},
\be
\int_{-\Lambda}^{\Lambda} \frac{\ud\omega}{\omega}\,
\rho_{12,12}(\omega,{\bp},T) \approx
\frac{1}{5}\,(e+p)\,
\left\langle {v_{\bp}^2}\right\rangle\,.
\la{eq:f-sumrule}
\ee
Here $\Lambda$ serves as a separation between 
the transport scale and the thermal scale.
For ultra-relativistic quasiparticles, as one might 
expect to find in the high-temperature gluon plasma, 
$\<v_{\bp}^2\>=1$. In that case,
comparison of (\ref{eq:f-sumrule}) 
with (\ref{eq:conti-sr}) reveals 
that at high temperatures, 
the spectral integral (\ref{eq:conti-sr})
receives an $\approx40\%$ 
contribution from the transport peak.
This means in particular that the prospect of 
determining the area under the transport peak 
from a numerically determined 
Euclidean correlator is realistic.


\acknowledgments{
I thank U.~Wiedemann and J.~Casalderrey-Solana for discussions at 
CERN, where this work was started.}

\bibliography{../../BIBLIO/viscobib.bib}

\end{document}